\documentclass{article}

\def\be{\begin{equation}}
\def\ee{\end{equation}}
\def\bea{\begin{eqnarray}}
\def\eea{\end{eqnarray}}
\def\({\left(}
\def\){\right)}
\def\[{\left[}
\def\]{\right]}
\def\<{\left<}
\def\>{\right>}

\begin{document}

\pagestyle{empty}
\vskip-10pt
\vskip-10pt
\hfill {\tt hep-th/0404150}
\begin{center}
\vskip 3truecm
{\Large\bf
Conformal anomaly of Wilson surface observables -- a field theoretical computation}\\ 
\vskip 2truecm
{\large\bf
Andreas Gustavsson\footnote{a.r.gustavsson@swipnet.se}
}\\
\vskip 1truecm
{\it Institute of Theoretical  Physics,
Chalmers University of Technology, \\
S-412 96 G\"{o}teborg, Sweden}\\
\vskip 5truemm
{\tt }
\end{center}
\vskip 2truecm
\noindent{{\bf Abstract:} We make an exact field theoretical computation of the conformal anomaly for two-dimensional submanifold observables. By including a scalar field in the definition for the Wilson surface, as appropriate for a spontaneously broken $A_1$ theory, we get a conformal anomaly which is such that $N$ times it is equal to the anomaly that was computed in hep-th/9901021 in the large $N$ limit and which relied on the AdS-CFT correspondence. We also show how the spherical surface observable can be expressed as a conformal anomaly.}
\newpage

\section{Introduction}
Wilson surface observables was introduced in \cite{Ga} in order to lift the loop equation formulation of Yang-Mills theory to a surface equation formulation of interacting six-dimensional $(2,0)$ supersymmetric gauge theories. Now the microscopic definition of non-abelian Wilson surfaces is not yet very well understood. We will therefore study smooth abelian Wilson surfaces, i.e. submanifolds, embedded in flat euclidean background. We can also consider the spontaneosly broken $A_1$ theory. The low energy theory should be described entirely in terms of the tensor multiplet fields, the massive degrees of freedom having decoupled. But the more interesting cases, from the point of view of the surface equation, of cusp singularities and self-intersections, will be postponed to future investigations. 

We will begin by studying the abelian Wilson surface for a selfdual two-form connection $B^+$, defined as
\be
W_0(\Sigma) = \exp {i\int_{\Sigma} B^+}.\label{Wilson0}
\ee
where we will assume that $\Sigma$ is a closed two-dimensional submanifold. As usual the vacuum expectation value $\<W(\Sigma)\>$ has to be renormalized. Closed submanifold observables of dimension $k$ have a conformal anomaly only if $k$ is even \cite{GW}. So in contrast to the Wilson loop (for which $k=1$), the surface observable (with $k=2$) that we will consider here will have an anomaly. Now this is true only for closed submanifold. If we have a boundary, we can get a contribution from boundary terms which can give rise to a conformal anomaly even for a Wilson line. An example of this is the non-zero difference between the renormalized straight Wilson line and the renormalized circular Wilson loop computed in \cite{DG}. Had there been no conformal anomaly these would have been equal since they are conformally equivalent, being related by an inversion map. Introducing a cut-off, thus cutting off the infinite line to a finite line, we get a contribution from the boundary, that thus contributes to a conformal anomaly even for $k=1$. 

This paper originated from the question of whether an analogous relation could hold between a Wilson surface being an infinitely extended plane, and a spherical Wilson surface observable. To answer this question we had to first compute the conformal anomaly exactly, to all orders in the conformal factor $\phi(x)$ (defined through the Weyl rescaling $G_{MN}(x)\rightarrow e^{2\phi(x)}G_{MN}(x)$ of the metric tensor) for the abelian Wilson surface. No exact computation of this anomaly had been done before. A computation had been done in \cite{HS},\cite{Gu}, but the result presented there is valid only for {\sl{constant}} Weyl rescalings, and is thus not directly applicable if one wants to study e.g. an inversion map. 

However this anomaly has been computed on the supergravity side for $A_{N-1}$ theory in the large $N$ limit in \cite{GW}. Here we will compute the corresponding anomaly for $N=2$ directly from a low-energy field theory, spontaneously broken by a large vacuum expectation value of one of the scalar fields. We will find an answer that is proportional to the anomaly that was obtained in \cite{GW} with the proportionality constant being equal to $N$. We will define the (euclidean) Wilson surface as
\be
W_1(\Sigma) =\exp {i\int_{\Sigma}d^2\sigma \(\frac{1}{2}\varepsilon^{\alpha\beta}\partial_{\alpha}X^M\partial_{\beta}X^N B^+_{MN}-\varphi(X)\sqrt{g(X)}\)}.\label{Wilson1}
\ee
where $\sigma^{\alpha}$ parametrizes $\Sigma$, and $\epsilon^{12}=1$. Here $\varphi$ is a scalar that is not to be confused with the one that we gave a large vacuum expectation value, and $d^2\sigma\sqrt{g(X)}$ is the invariant measure on $\Sigma$. The scalar $\varphi$ has vacuum expectation value zero and describes fluctations about the vacuum expectation value. We will motivate this definition in section \ref{ph}.

The redaction of this paper is as follows. In section \ref{r} we discuss our regularization. In section \ref{n} we determine the normalization of the action. In section \ref{p} we study the spherical Wilson surface, which we can compute explicitly. In sections \ref{c} and \ref{m} we make an exact computation of the conformal anomaly for $\<W_0\>$ and $\<W_1\>$ respectively, that is, for any Weyl rescaling in any background. In section \ref{s} we show that the renormalized spherical Wilson surface is nothing but a certain conformal anomaly. Comparing this with the explicit results in section \ref{p} will serve as a consistency check of our results. In section \ref{ph} we justify our definition of $W_1$ by making contact with Wilson loops in four dimensional gauge theory.

\section{The regularization}\label{r}
We start by considering the free abelian theory. The vacuum expectation value of the corresponding Wilson surface (\ref{Wilson0}) is given by
\be
\<W_0(\Sigma)\> = \exp -I_0(\Sigma)
\ee
where
\be
I_0(\Sigma) = \int_{(X,Y)\in \Sigma\times \Sigma} \Delta(X,Y).
\ee
Here $\Delta$ denotes the bi-two-form gauge field propagator,
\be
\Delta(X,Y) = \frac{1}{4}dX^M\wedge dX^N\otimes dY^P\wedge dY^Q \<B^+_{MN}(X)B^+_{PQ}(Y)\>
\ee

We have to regularize $I(\Sigma)$. Since we want to compute a conformal anomaly we should use a regularization that does not break covariance and gauge invariance of the Wilson surface. Apriori such a regularization should exist given that there are no other anomalies than the conformal anomaly present. We will start by considering flat backgrounds.\footnote{In a curved background there is generically no symmetry between the directions in which we could move one copy of the surface in order to regularize it. It therefore matters which direction we choose. But such a direction can not be written in a coordinate independent way, unless of course the space has some isometry. But even if we restricted to spaces which has some isometry, when we make a Weyl transformation that isometry will generically be destroyed. There should be some way to overcome this problem since a covariant and gauge invariant regularization should exist. Maybe such a regularization would be to sum over all directions. We must also find a way to define the direction at a different point given that we have picked a direction (i.e. an orthonormal vector) at some point. Just parallel transporting this vector along the surface may bring it outside the three allowed directions. But if we project the parallel transported vector to this three dimensional space, and also normalize it to unit length, then we have constructed a smooth normalized vector field $\alpha^M(X)$ over the whole surface that only depends (in a unique way) on one direction $\alpha^M(\bar{X})$ at some point $\bar{X}$ on the surface. Choosing normal coordinates at any point, say $\bar{X}$, we have the metric tensor $\delta_{MN}$ at that point. We now integrate over all allowed directions $\alpha^M$. which constitutes an $S^1$. (If there are more possible directions than those which lie in an $S^1$ the problem happens to have more symmetries than we asked for. Then one can restrict to any of the possible $S^1$'s and which one one chooses will not matter.) This now seems to be a coordinate independent regularization of our submanifold observable that may work in any smoothly curved background.} This will actually not to be such a serious restriction as it may seem at first sight. Indeed we will in the end of section \ref{c} show that in order to get the conformal anomaly in a generic background, it is sufficient to compute it for the following two separate cases
\begin{enumerate}
\item{a generically curved background with constant conformal factor (i.e. Weyl rescalings that are constant rescalings of the metric).} 
\item{a flat background with conformal factor such that the Weyl transformed metric is still flat.}
\end{enumerate}
Assume now that the six dimensional background metric is $G_{MN}(x) = \delta_{MN}$, and that the surface $\Sigma$ is confined to a hyperplane, say $x^0 = X^0$. We then regularize $I(\Sigma)$ by translating one copy of $\Sigma$ to a new position $x^0 = Y^0 = X^0 + \epsilon$. This regularization prescription can of course be reformulated in any coordinate system. If we have a flat space with metric tensor of the form $G_{MN}(x) = e^{2\phi(x)}\delta_{MN}$ we can always make a diffeomorphism $x^M \mapsto \tilde{x}^M$ to a normal coordinate system $\tilde{x}^M$, which brings the metric tensor to $\tilde{G}_{MN}(\tilde{x}) = \delta_{MN}$. We must now assume that the surface is located at some point in some direction, that we will denote $\alpha^M$, when it is expressed in terms of a {\sl{normal}} coordinate system.\footnote{Since we are interested in the change under a conformal coordinate transformation we also have to assume that the surface is located at a point in the origial coordinates as well. This can be ashieved if we assume that there are two directions, say $x^0$ and $x^5$ along which the surface is at a point.} We translate one copy of $\Sigma$ a distance $\epsilon$ in a direction $\alpha^M$. In terms of the original coordinate system $x^M$ this regularization reads (noting that straight lines are geodesics in flat space and that geodesics are mapped to geodesics under diffeomorphisms):

\noindent{Displace all points on $\Sigma$ a certain distance $\epsilon$ along geodesics that intersect $\Sigma$ orthogonally and emanate from it in the $\alpha^M$-direction, where $\alpha^M$ is a unit vector such that the surface lies at a point in the direction of this vector when expressed in terms of normal coordinates.} 

\noindent{These geodesics will automatically also intersect displaced surface $\Sigma'$ orthogonally.}

\section{Normalization and selfduality}\label{n}
In order for the Wilson surface as we have defined it to be well-behaved we should normalize the gauge field so that the bosonic part for the free $(2,0)$ supersymmetric action is (after Wick rotation)\footnote{This normalization also follows from requiring holomorphic factorization} \cite{Henning},\cite{Gustav}
\be
S=\frac{1}{8\pi}\int\(H\wedge*H+2d\phi^a\wedge*d\phi^a\),
\ee
where $H=dB$ is the non-chiral field strength, and $\phi^a$ denote the five Lorentz scalars in the $(2,0)$ tensor multiplet. The factor $2$ in front of the scalar kinetic term follows from requiring the supercharges derived from this action to reduce to the supercharges in four-dimensional $N=4$ supersymmetric gauge theory. This is appropriate because we have defined our Wilson surface in such a way that it reduce to the Wilson loop in such a four dimensional supersymmetric theory. We have thus no freedom to change any of our normalization constants, once we have  defined the Wilson surface, Eq. (\ref{Wilson1}). We defer the details to section \ref{ph}.

In this normalization we get the gauge field propagator $\Delta$ in Feynman gauge and the scalar field propagator $D$ respectively as
\bea
\Delta(\tilde{y},\tilde{x})&=&\frac{1}{\pi^2}d\tilde{y}^M\wedge d\tilde{y}^N\otimes d\tilde{x}_M\wedge d\tilde{x}_N |\tilde{y}-\tilde{x}|^{-4}\cr
D^{ab}(\tilde{y},\tilde{x})&=&\delta^{ab}\frac{1}{\pi^2}|\tilde{y}-\tilde{x}|^{-4}
\eea
in normal coordinates $\tilde{x}^M$. If the background is curved there will be $O(|\tilde{y}-\tilde{x}|^{-2})$ corrections to the short-distance behaviour of these propagators. See \cite{HS} and Eq. (\ref{short}).

If we in euclidean signature separate $H$ into selfdual and anti-selfdual parts as $H=H^+ + H^-$ we have
\be
\<H^+(x)H^+(y)\>=\frac{1}{2}\(\<H(x)H(y)\>+i\<H(x)*H(y)\>\),
\ee
where $*$ denotes the Hodge duality operator. Correspondingly we have
\be
\<B^+(x)B^+(y)\>=\frac{1}{2}\(\<B(x)B(y)\>+i({\mbox{something real}})\).
\ee
In this letter we will not care about the phase factor of the selfdual Wilson loop expectation value. We will thus simply divide the ordinary gauge field propagator by $2$ and ignore the imaginary part. We should also not forget a factor $2$ that comes from the fact that $dx^M\wedge dx^N\otimes dy_M\wedge dy_N=2\sum_{M<N}dx^M\wedge dx^N\otimes dy_M\wedge dy_N$.

\section{A preliminary example -- the spherical Wilson surface}\label{p}
We can illustrate our regularization by taking the surface observable for a 2-sphere $S^2_{R,0}$ of radius $R$ centered at the origin and located at $\tilde{x}^0=\tilde{x}^4=\tilde{x}^5=0$. The selfdual regularized spherical Wilson surface is 
\be
I_0(\epsilon)=\frac{1}{2\pi^2}\int_{S^2_{R,0}} d\tilde{X}^M\wedge d\tilde{X}^N\int_{S_{R,\epsilon}^2} d\tilde{Y}_M\wedge d\tilde{Y}_N \frac{1}{|\tilde{X}-\tilde{Y}|^4}\label{res}
\ee
with $(\tilde{Y} - \tilde{X})^M \alpha_M = \epsilon$. It is easy to compute the sphere explicitly by fixing a point $\bar{X}$ on the north pole, using the rotational symmetry. We get the result
\be
I_0(\epsilon)=4\(\frac{R^2}{\epsilon^2}-\ln\(\frac{2R}{\epsilon}\)+\frac{1}{4}+O(\epsilon)\).\label{sphere}
\ee
Computing the similar thing for the $A_1$ model we get
\be
I_1(\epsilon)=4\(\frac{2R^2}{\epsilon^2}-\ln\(\frac{2R}{\epsilon}\)+O(\epsilon)\).\label{sphere1}
\ee
We will rederive these results (or rather the corresponding renormalized results) in section $5$ as conformal anomalies of a plane. 

The spherical Wilson surface has also been computed on the supergravity side in the large $N$ limit in \cite{BCFM} with the result
\be
I_{\infty}(\epsilon)=4N\(\frac{R^2}{\epsilon^2}-\ln\(\frac{2R}{\epsilon}\)-\frac{1}{2}+O(\epsilon)\).\label{spherei}
\ee

We now make the observation that if we as original coordinates $x^M$ take
\be
\tilde{x} = x/|x|^2
\ee
the equation for the sphere is given by $|X|=1/R$. Since we do not change the metric under this transformation, the radius of the sphere changes to $1/R$. If we instead make the rescaling
\be
\tilde{x}^M = R^2 x^M,
\ee
the new sphere is given by the equation $|X|=1/R$. We see that the new sphere is given by $|X| = 1/R$, irrespectively of whether $x$ was related to the normal coordinates via an inversion through the center of the sphere, or via a certain rescaling. But in both cases we end up with the same regularized result (\ref{res}) according to our regularization prescription. It follows that the conformal anomaly must also be the same for these two transformations.
 
Instead of transforming the coordinates and keeping the metric fixed, we can keep the coordinates fixed and Weyl rescale the metric. For the inversion map we get the Weyl rescaling
\be
\delta_{MN} \rightarrow |x|^{-4}\delta_{MN},
\ee
and the corresponding conformal factor $\phi = -2\ln|x|$.

To be slightly more general we can consider the effect of this inversion map on a sphere $S^2_{R,a}$ centered a distance $a<R$ from the origin. This maps the sphere to a new sphere of radius 
\be
\tilde{R} =\frac{R}{R^2-a^2}.
\ee
The anomaly that corresponds to this rescaling of the radius can be read off from (\ref{sphere}) as
\be
I_{0}(\tilde{R})-I_{0}(R) = 4\(\ln R-\ln \tilde{R}\) = 4 \ln(R^2-a^2).
\ee
In flat space the conformal anomaly can only be a linear combination of covariant terms that can be build up of the induced metric $g_{\alpha\beta}$ and the second fundamental form $\Omega^M_{\alpha\beta}$ on the surface. We will use the abbreviation $\Omega^M = g^{\alpha\beta}\Omega^M_{\alpha\beta}$ and let $R_{(2)}$ denote the Ricci scalar that is computed from $g_{\alpha\beta}$. For infinitesimal conformal transformations, the descent equations reqire the conformal anomaly to be conformally invariant. The only combinations that do not involve derivatives of the conformal factor are $\(|\Omega|^2-4g^{\alpha\beta}P_{\alpha\beta}\)\phi$ and $R_{(2)}\phi$, as in \cite{HS} using the same notations as there. In flat space and for a sphere both these terms are a constant times $\phi/R^2$. In this case the anomaly can only be a certain linear combination (with constant, i.e. numerical, coefficients) of the following terms,
\bea
A&=&\int_{\S^2} d^2\sigma \sqrt{g} \frac{4}{R^2}\phi\cr
B&=&\int_{\S^2} d^2\sigma \sqrt{g} \Omega^M\partial_M\phi\cr
C&=&\int_{\S^2} d^2\sigma \sqrt{g} g^{\alpha\beta}\partial_{\alpha}\phi\partial_{\beta}\phi\cr
D&=&\int_{\S^2} d^2\sigma \sqrt{g} \partial_M\phi\partial_M\phi.
\eea
No terms of order higher than quadratic in $\phi$ can occur as will be clear later. In flat space, we can rewrite all higher derivatives of $\phi$ in terms of powers of terms with only one derivative of $\phi$ (as can be deduced from Eq (\ref{flat}). 

For the case that $\Sigma = S^2_{R,a}$ we have $\Omega^n_{ij}=-\delta_{ij}/R$. For inversion, $\phi=-2\ln|x|$, of this sphere we get
\bea
A&=&-8\pi\(2\ln(R^2-a^2)+2\frac{R^2+a^2}{2Ra}\ln\frac{R+a}{R-a}-4\)\cr
D-B=C&=&8\pi\(\frac{R^2+a^2}{2Ra}\ln\frac{R+a}{R-a}-2\)
\eea
so what we see now is that any combination of the form 
\be
\frac{1}{4\pi}\[-A + f\cdot(B-D) + (f-2)C\]\label{form}
\ee
yields the answer $4\ln(R^2-a^2)$. So from this example alone, we could not determine the conformal anomaly uniquely, but only up to some number $f$. But we should notice that Eq (\ref{form}) is a very general result since the observation that inversion and rescaling of a sphere yield the same result should hold in any theory, interacting or not, that contains surface observables, such as the $(2,0)$ supersymmetric $A_N$ theories.

\section{The conformal anomaly}\label{c}
We will first assume that the Riemann curvature tensor vanishes, both before as well as after a conformal transformation. After a conformal transformation, $\delta_{MN}\rightarrow e^{2\phi(x)}\delta_{MN}$, the Riemann curvature tensor is given by
\be
R_{MPNQ} = e^{2\phi}\delta_{MN}\(\partial_P\partial_Q\phi - \partial_P\phi\partial_Q\phi + \frac{1}{2}\delta_{PQ} |\partial \phi|^2\)\pm {\mbox{permutations}},
\ee
where we use the abbreviation $|\partial\phi|^2\equiv \delta^{MN}\partial_M\phi\partial_N\phi$. Contracting indices one finds that $R_{MPQN}=0$ if and only if
\bea
\partial^2 \phi + 2|\partial \phi|^2 &=& 0\cr
\partial_M\phi\partial_N\phi - \partial_M\partial_N \phi - \frac{1}{2}\delta_{MN}|\partial \phi|^2 &=& 0.\label{flat}
\eea
Requiring the space to be flat, we can find a Riemann normal coordinate system $\tilde{x}^M$ around a point with coordinate $X = 0$ in which the metric is $(ds)^2 = d\tilde{x}^M d\tilde{x}^M$. A normal coordinate system may be constructed by noticing that geodesics in this coordinate system are straight lines. In a any conformally flat space we have the geodesic differential equation
\be
\ddot{X}^M - |\dot{X}|^2 \partial^M \phi + 2\dot{X}^M\dot{X}^N\partial_N\phi = 0.\label{diff}
\ee
where $\delta_{MN}$ is used to contract indices here (though the metric is $e^{2\phi(x)}\delta_{MN})$. The dot denotes a derivative with respect to a parameter $\tau\in [0,1]$ that parametrizes the geodesic. Contracting the geodesic equation with $\dot{X}_M$ we get
\be
\frac{d}{d\tau}\(\ln |\dot{X}| + \phi\) = 0,
\ee
which we can integrate to get 
\be
|\dot{X}(\tau)| = \epsilon e^{-\phi(X(\tau))},
\ee
Assuming the boundary condition
\be
\dot{X}^M(0) = \tilde{\epsilon}\alpha^M
\ee
where $\tilde{\epsilon} \equiv \epsilon e^{-\phi(0)}$, and $\alpha$ a unit vector, we can solve (\ref{diff}) iteratively. If we use the defining property of a normal coordinate system that geodesics are straight lines,
\be
\tilde{x}^M = \tilde\epsilon \alpha^M
\ee
we get the diffeomorphism between the original and the normal coordinate systems,
\bea
x^M &=& \tilde{x}^M + \frac{1}{2}|\tilde{x}|^2 I^{MN}(\tilde{x}) \partial_N \phi \cr
&+& \frac{1}{6}(|\tilde{x}|^2 2\tilde{x}^N I^{NP}(\tilde{x})\partial_P\phi+|\tilde{x}|^2\tilde{x}^N\partial_N\partial^M\phi-2|\tilde{x}|^2\tilde{x}^PI^{MN}(\tilde{x})\partial_N\phi\partial_P\phi\cr
&&-2|\tilde{x}|^2\tilde{x}^MI^{NP}(\tilde{x})\partial_P\phi\partial_N\phi-2\tilde{x}^M\tilde{x}^N\tilde{x}P\partial_P\partial_N\phi) + {\cal{O}}(|\tilde{x}|^4),
\eea
where $I^{MN}(\tilde{x})\equiv \delta^{MN}-2\frac{\tilde{x}^M\tilde{x}^N}{|\tilde{x}|^2}$.

We must now find some direction $\alpha$ in which the surface $\Sigma$ is located at a point in terms of the normal coordinate system. According to our regularization prescription, it is along this direction we should translate one copy of the surface a cut-off distance $\epsilon$. We will assume that $\Sigma$ is located at $X^0 = X^5 = 0$ and make the ansatz $\alpha = (\alpha_1,0,0,0,0,\alpha_5)$. We thus look for a hyperplane in which $\Sigma$ is confined, of the form 
\be
\alpha_M \tilde{X}^M = 0.\label{zero}
\ee
We rewrite the hyperplane $X^0 = X^5 = 0$ in terms of normal coordinates and contract with $\alpha$ (using (\ref{zero})) to find the condition
\be
\alpha^M\partial_M\phi(0) = 0.
\ee
We thus find a direction
\be
\alpha = \frac{(\partial_5 \phi,0,0,0,0,-\partial_0\phi)}{\sqrt{(\partial_0\phi)^2 + (\partial_5\phi)^2}}.
\ee

We must now compute the coordinate $Y$ to which a point $X = 0$ on $\Sigma$ is translated along this geodesic. We get
\be
Y^M = \tilde{\epsilon}\alpha^M + \frac{1}{2} \tilde{\epsilon}^2 \partial^M\phi + \frac{1}{6}\tilde{\epsilon}^3 \(\alpha^P\partial_P \partial^M\phi - 2\alpha^M |\partial\phi|^2 - 2\alpha^M\alpha^N\alpha^P\partial_P\partial_N\phi\) + {\cal{O}}\(\tilde{\epsilon}^4\)
\ee
which when restricting to flat space (\ref{flat}) reduces to
\be
Y^M = \tilde{\epsilon}\alpha^M + \frac{1}{2} \tilde{\epsilon}^2 \partial^M\phi - \frac{1}{4}\tilde{\epsilon}^3 \alpha^M |\partial\phi|^2 + {\cal{O}}\(\tilde{\epsilon}^4\)
\ee
Each point $Y$ on the displaced surface $\Sigma'$ corresponds to a point $\bar{X}$ on $\Sigma$ via this displacement along geodesics, 
\be
\Sigma \ni \bar{X} \mapsto Y = Y(\bar{X})\in \Sigma'.
\ee
We can then write $I_0(\epsilon)$ as an integral over $\Sigma\times\Sigma$ as follows,
\bea
I_0(\epsilon) &=& \int_{\Sigma}dX^M\wedge dX^N \int_{\Sigma'}dY_M\wedge dY_N \frac{1}{|X-Y|^4}\cr
&=&\int_{\Sigma}dX^M\wedge dX^N\int_{\Sigma}d\bar{X}^R\wedge d\bar{X}^S \frac{\partial Y^M}{\partial \bar{X}^R}\frac{\partial Y^N}{\partial \bar{X}^S}\frac{1}{|X-Y(\bar{X})|^4}
\eea
We can compute this integral by first performing the integral over $\bar{X}$ while keeping $X$ fixed. But we can just as well keep $\bar{X}$ fixed and perform the integral over $X$ first. The latter alternative will be more convenient since we have already expanded everything about $\bar{X}$, so we will choose that alternative.

For later reference we compute
\be
|X-Y(\bar{X})|^2 = |X-\bar{X}|^2 + \tilde{\epsilon}^2(1-(X-\bar{X})^M\partial_M\phi)+\tilde{\epsilon}^4\(-\frac{1}{4}|\partial \phi|^2+(X-\bar{X})^M(...)_M\) + {\cal{O}}\(\tilde{\epsilon}^5\)\label{difference}
\ee
and
\be
dY^{1}\wedge dY^{2} = d\bar{X}^1\wedge d\bar{X}^2 \(1 + \tilde{\epsilon}^2\frac{1}{2}\delta^{\alpha\beta}\(\partial_{\alpha}\partial_{\beta}\phi - 2\partial_{\alpha}\phi\partial_{\beta}\phi\) + O(\tilde{\epsilon}^3)\)
\ee
which, using the flatness conditions Eq (\ref{flat}), becomes
\be
dY^{1}\wedge dY^{2} = d\bar{X}^1\wedge d\bar{X}^2 \(1 + \tilde{\epsilon}^2\frac{1}{2}\(-|\partial_{\alpha}\phi|^2 - |\partial\phi|^2\) + O(\tilde{\epsilon}^3)\).\label{change}
\ee
The conformal anomaly is given by the constant (i.e. $\epsilon$-independent) terms in the difference
\be
I(\epsilon)[e^{2\phi}G_{MN},\Sigma] - I(\epsilon)[G_{MN},\Sigma].
\ee
In general there is probably no simple way to compute any of these two terms separately, but the difference can be computed - even for finite conformal transformations - and will be given by a local expression integrated over $\Sigma$.

The trick is to rewrite this difference as the sum of the contribution of a local (both $X$ as well as $\bar{X}$-dependent) rescaling of $\tilde{\epsilon}(\bar{X})^2$ to something which is of the form $\tilde{\epsilon}(\bar{X})^2(1-\xi(X,\bar{X}))+\tilde{\epsilon}(\bar{X})^4\kappa(X,\bar{X})+O(\tilde{\epsilon}^5)$ (to be read off from Eq (\ref{difference})) at each fixed $\bar{X}$, plus the contribution of the $X$-independent rescaling of $\tilde\epsilon(\bar{X})$ to $\epsilon$. Throughout all the computation we will let $\bar{X}$ be a fixed point on $\Sigma$.
The idea is thus to recast the variation in a form that implies a (local) rescaling of $\epsilon$ for each fixed $\bar{X}$ because then, upon Taylor expanding about $\tilde{\epsilon}^2$, the anomaly becomes a certain expression which involve certain derivatives with respect to $\tilde\epsilon^2$ (of the form $\tilde{\epsilon}^{2p}(\partial/\partial{\tilde{\epsilon}^2})^q$ for $p,q=1,2,...$) of integrals that will diverge as $\tilde{\epsilon}\rightarrow 0$ (see Eq (\ref{derivatives})). Pulling these derivatives outside the integrals, they will kill certain divergences in integrals whose divergent part we can compute. This is a straightforward generalization of the method used in \cite{Gu}. In this reference all the divergent integrals that we will need here can also be found. 

We should thus recast the Weyl transformed Wilson surface in the form (dropping the $\int_{\Sigma}d\bar{X}\wedge d\bar{X}$-integration for the time being)
\be
\int_{\Sigma}\Delta(\tilde{\epsilon}^2(1-\xi) + \tilde{\epsilon}^4 \kappa + O(\tilde{\epsilon}^5))
\ee
Explicitly we read off from Eq (\ref{difference}) that
\bea
\xi &=& (X-\bar{X})^M\partial_M\phi\cr
\kappa &=& -\frac{1}{4}|\partial\phi|^2-(X-\bar{X})^M(...)_M.
\eea
We then Taylor expand in $\alpha \equiv \tilde{\epsilon}^4 \kappa + O(\tilde{\epsilon}^5)$ about $\tilde{\epsilon}^2$ as
\bea
&&\int \(\Delta(\tilde{\epsilon}^2) - \xi\tilde{\epsilon}^2 \Delta'(\tilde{\epsilon}^2) + \frac{1}{2!}(\xi\tilde{\epsilon}^2)^2 \Delta''(\tilde{\epsilon}^2) - ...\)\cr
&&+\int \(\alpha \Delta'(\tilde{\epsilon}^2) - \alpha \xi \tilde{\epsilon}^2 \Delta''(\tilde{\epsilon}^2) + \frac{1}{2!} \alpha (\xi\tilde{\epsilon}^2)^2 \Delta'''(\tilde{\epsilon}^2) - ...\)\cr
&&+\int \frac{1}{2!}\(\alpha^2 \Delta''(\tilde{\epsilon}^2) - \alpha^2\xi\tilde{\epsilon}^2 \Delta'''( \tilde{\epsilon}^2) + \frac{1}{2!} \alpha^2 (\xi\tilde{\epsilon}^2)^2 \Delta''''(\tilde{\epsilon}^2) - ...\)+...\label{derivatives}
\eea
The $+...$ either contain terms of higher order in $\tilde{\epsilon}$ or in $X-\bar{X}$ and will not be sufficiently divergent to give any finite contribtion to the conformal anomaly. We will explain this in detail at the end of this section.

Integrating $\Delta$, or $\Delta$ multiplied by any combination of $\xi, \eta$ and $\kappa$, yields only two types of divergences, $\tilde{\epsilon}^{-2}$ and $\ln \(\tilde{\epsilon}^2\)$. Finite terms arise from quadratic divergences in 
\be
\int\(-\kappa -2\xi\kappa - 3\xi^2\kappa\) \Delta(\tilde{\epsilon}^2) 
\ee
as well as from the logarithmic divergences in
\be
\int\(-\xi-\frac{1}{2}\xi^2\)\Delta(\tilde{\epsilon}^2).
\ee
From the quadratic divergence coming from $\(-\kappa+...\) \Delta$ we get the finite contribution (the derivative kills the divergence)
\be
\frac{1}{4\pi}|\partial\phi|^2.
\ee
Similarly the logarithmic divergence $\ln(\tilde{\epsilon}^2)$ in $\(-\xi-\frac{1}{2}\xi^2\)\Delta$ gives the finite contribution
\be
\frac{1}{4\pi}\(\Omega^M\partial_M \phi + |\partial_{\alpha}\phi|^2\).
\ee
The change of variables produces another $|\partial_{\alpha}\phi|^2$, coming from a quadratic divergence (see Eq (\ref{change})), and gives rise to the finite contribution (the $\epsilon^2$ in Eq (\ref{change}) cancelling the $\epsilon^{-2}$-divergence)
\be
\frac{1}{4\pi}\(-|\partial_{\alpha}\phi|^2 - |\partial\phi|^2\)
\ee
These are all the finite contributions there is. Summing them up, we have 
\be
\frac{1}{4\pi}\(-\(\frac{3}{2}\Omega^2+R_{(2)}\)\phi + \Omega^M\partial_M \phi -|\partial_{\alpha}\phi|^2 - |\partial\phi|^2\).
\ee
where the piece $\(\frac{3}{2}\Omega^2+R_{(2)}\)\phi$ was computed in \cite{HS},\cite{Gu} and arises entirely from the rescaling of $\epsilon$ from $\tilde{\epsilon}$. Now this result is exact (to all orders in $\phi$). We also notice that it satisfies the general constraint Eq (\ref{form}) for a sphere if we choose $f=1$.

The derivative terms of $\phi$ derived above in flat space, must  carry over unchanged (just covariantized in the obvious way) to curved space. This is so for the following two reasons:
\begin{enumerate}
\item{There can be no terms of higher than quadratic orders in $\partial\phi$ because, by dimensional reasons each such term must be accompanied by $X-\bar{X}$ or by $\epsilon$. Now since there are no higher divergences than $\epsilon^{-2}$, terms of order $O(|X-\bar{X}|^3)$ or $O(\epsilon^3)$ or similar, will give only $O(\epsilon)$-terms.}
\item{There can not be any terms involving more than one derivative of $\phi$. Such terms would have been of the form $D\partial\phi$ where $D$ is a covariant derivative. But apparently no Christoffel symbol $\Gamma_{\alpha\beta}^{\gamma}$ is present in the flat space result above. To rule out the possibility of $\Gamma_{MN}^P$ is a bit harder though. This is zero for the case we computed on above. But the Wess-Zumino consistency condition requires the conformal anomaly for infinitesimal transformations to be conformally invariant. But $D\partial\phi$ is not conformally invariant. Such a term is therefore not allowed in the anomaly.}
\end{enumerate}

In a curved background the propagator gets modified. But that modification will only affect the piece $(...)\phi$ in the anomaly. This is so, because these terms in the propagator are not sufficiently divergent as $X$ approaches $Y$, to give any finite contribution to terms that involves derivatives of $\phi$. But for $\phi$ constant we can safely use the regularization of displacing the surface in some direction as far as the conformal anomaly concerns, because then we can choose normal coordinates about some point on the surface, such that the coordinate axes become geodesics, and we can displace along some suitable coordinate axis (along which the surface is at a point). If now $\phi$ is constant, we can use the same normal coordinate system after the Weyl transformation, and the difference will not depend on the detailed curved along which we displaced the surface because for constant $\phi$ only the rescaling of $\epsilon\rightarrow \tilde{\epsilon}=\epsilon e^{-\phi}$ matters. This way we recover the same expression as was first obtained in \cite{HS}, as long as $\phi$ is a constant.

Generically a two-dimensional surface will of course not be confined to any hyperplane when expressed in terms of normal coordinates. But, given a point $X$ on the surface, we can always find a two-dimensional hyperplane in which the surface is confined up to cubic order in the distance along the surface. That will be good enough for our purposes of computing the conformal anomaly, the cubic term giving no finite contribution to the anomaly.

\subsection{The $A_1$ model}\label{m}
A conformally invariant action for a scalar field in six dimensions is
\be
\frac{1}{4\pi}\int d^6 x \sqrt{G}\(G^{MN}\partial_M\varphi\partial_N\varphi-\frac{1}{5}R\varphi^2\)
\ee
where $R$ is the Ricci scalar of the background. Expanding in normal coordinates about $x$ and using that the metric in normal coordinates is 
\be
G_{MN}(y)=\delta_{MN}+\frac{1}{3}R_{MPNQ}(x)(y-x)^P(y-x)^Q+...
\ee
we get after some computations, similar to the computation of the gauge field propagator which was carried out in \cite{HS}), that the scalar field propagator is given by
\be
D(y,x) = \frac{1}{\pi^2}|y-x|^{-4}\(1-\frac{1}{3}P_{MN}(x)(y-x)^{M}(y-x)^{N}+O(|y-x|^3)\),\label{short}
\ee
where $P_{MN}=\frac{1}{4}\(R_{MN}-\frac{1}{10}RG_{MN}\)$. We notice that there is no $R$ dependence left in this propagator (although $R$ appears in the action). This have cancelled out through a lucky numerical coincidence. Of course this had to happen since the conformal anomaly can not contain any term proportional to $R$. 

We define the Wilson surface observable in our spontaneosly broken $A_1$ theory as
\be
W_1(\Sigma)=\exp i\int_{\Sigma}d^2\sigma \(\partial_1X^M\partial_2X^N B^+_{MN}-\varphi(X)\sqrt{g(X)}\)
\ee
where $\phi$ is a scalar which has no vacuum expactation value. We will justify this definition in section \ref{ph}. 

We should thus consider
\be
\int d^2\sigma \int d^2\bar{\sigma} \sqrt{g(X(\sigma))g(Y(\bar{\sigma}))}D(X(\sigma),Y(X(\bar{\sigma})))
\ee
The induced metric at $Y^M=\bar{X}^M+\epsilon e^{-\phi}\alpha^M+...$ on the displaced surface is
\bea
g_{\alpha\beta}(Y)&=&\partial_{\alpha}Y_M\partial_{\beta}Y_M\cr
&=&\partial_{\alpha}\bar{X}^P\partial_{\beta}\bar{X}^Q\partial_PY^M(\bar{X})\partial_QY^M(\bar{X})
\eea
and, noting that $\partial_{\alpha}X^P\alpha_P = 0 = X^P\partial_{\alpha}\alpha_P$, we get the metric on the displaced surface, in normal coordinates about $\bar{X}$, as
\be
g_{\alpha\beta}(Y)=g_{\alpha\beta}(\bar{X})\(1-\frac{1}{2}\tilde{\epsilon}^2|\partial\phi|^2+O(\tilde{\epsilon}^3)\)+... .
\ee
Furthermore, still in  normal coordinates about $\bar{X}$ (with respect to the background), we get for the measure on $\Sigma$,
\be
\sqrt{g(X)} = 1 + \frac{1}{2}(X-\bar{X})^{\alpha}(X-\bar{X})^{\beta}g^{\gamma\delta}\(\Omega_{\alpha\gamma}\cdot\Omega_{\beta\delta}+\frac{1}{3}R_{\alpha\gamma\beta\delta}\)+... .
\ee
We will make use of the relation $R_{\alpha\gamma\beta\delta}=W_{\alpha\gamma\beta\delta}+4g_{{}^[\alpha_[\beta}P_{\delta_]\gamma^]}$.
Using the general formalism developed in section \ref{c} and results from \cite{Gu}, we get the anomaly from the scalar as the sum of the contribution from the propagator which is
\be
\frac{1}{\pi^2}\[-\(\frac{\pi}{8}\(|\Omega|^2+2g^{\alpha\beta}g^{\gamma\delta}\Omega_{\alpha\gamma}\cdot\Omega_{\beta\delta}\)+\frac{\pi}{3}g^{\alpha\beta}P_{\alpha\beta}\)\phi + \frac{\pi}{4}\(\Omega^M\partial_M\phi+|\partial\phi|^2+|\partial_{\alpha}\phi|^2\)\],
\ee
from the measure,
\be
-\frac{1}{\pi^2}\(-\frac{\pi}{2}g^{\alpha\beta}g^{\gamma\delta}\Omega_{\alpha\gamma}\cdot\Omega_{\beta\delta}-\frac{\pi}{6}\(g^{\alpha\beta}g^{\gamma\delta}W_{\alpha\gamma\beta\delta}+2g^{\alpha\beta}P_{\alpha\beta}\)\)\phi
\ee
and from the change of variables from $Y$ to $\bar{X}$ which produces a finite piece when multiplying the quadratic divergence, which is
\be
-\frac{1}{2\pi}|\partial\phi|^2
\ee
Using the identity
\be
g^{\alpha\beta}g^{\gamma\delta}\Omega_{\alpha\gamma}\cdot\Omega_{\beta\delta}=|\Omega|^2+R_{(2)}-g^{\alpha\beta}g^{\gamma\delta}W_{\alpha\gamma\beta\delta}-2g^{\alpha\beta}P_{\alpha\beta}
\ee
we get the sum of all these contributions as
\be
\frac{1}{4\pi}\[\(\frac{1}{2}\(|\Omega|^2-4g^{\alpha\beta}P_{\alpha\beta}\)+R_{(2)}-\frac{1}{3}g^{\alpha\beta}g^{\gamma\delta}W_{\alpha\gamma\beta\delta}\)\phi +
\Omega^M\partial_M\phi-|\partial\phi|^2+|\partial_{\alpha}\phi|^2\]
\ee
We may verify that this form of the anomaly passes the test that, in a flat background, rescaling and inversion of a sphere should yield the same answer. Indeed for the sphere we get no contribution from the scalars to the anomaly. 
 
Adding this to the contribution from the gauge field, which in curved background is\footnote{An error of a factor $2$ in the last term in the gauge field propagator as computed in \cite{HS} ($...-\frac{1}{3}(X-X')(X-X')W...$ in Eq. (17) in \cite{HS} should be $...-\frac{2}{3}(X-X')(X-X')W...$ which can be seen by e.g.considering a Ricci flat background) also propagated to \cite{Gu} which in those papers led to wrong sign for the term which involves the Weyl tensor in this anomaly, which has been corrected here.}
\be
\frac{1}{4\pi}\[-\(\frac{3}{2}\(|\Omega|^2-4g^{\alpha\beta}P_{\alpha\beta}\)+R_{(2)}-\frac{1}{3}g^{\alpha\beta}g^{\gamma\delta}W_{\alpha\gamma\beta\delta}\)\phi + \Omega^M\partial_M \phi -|\partial_{\alpha}\phi|^2 - |\partial\phi|^2\]
\ee
we get
\be
\frac{1}{4\pi}\[-\(|\Omega|^2-4g^{\alpha\beta}P_{\alpha\beta}\)\phi + 2\Omega^M\partial_M\phi - 2|\partial\phi|^2\]\label{anom}
\ee
which, when covariantly integrated over $\Sigma$, indeed is proportional to the result obtained in \cite{GW}, with proportionality constant equal to $N$.

\section{The spherical Wilson surface as a conformal anomaly}\label{s}
The bi-two-form gauge field propagator $\<B(x)B(y)\>$ can at most change by a gauge transformation under a conformal transformation since it is only the gauge fixing term which is not conformally invariant. But we can also show this explicitly by considering the field strength propagator $\<H(x)H(y)\>$ where $H\equiv dB$. The field strength is a primary field of weight $3$ under the conformal group,
\be
H_{MNP}(x) \rightarrow H'_{MNP}(x') = \Omega(x)^3 H_{RST}D^{RST}{}_{MNP}({\cal{R}}(x)).
\ee
Here ${\cal{R}}_{M}{}^N(x)=\Omega(x)\frac{\partial x'_M}{\partial x_N}$ where $\Omega$ is such that this is an $x$-dependent $O(6)$ rotation, $D$ is a representation matrix and $I_{MN}(x)=\delta_{MN}-2\frac{x_Mx_N}{|x|^2}$. According to the general formalism developed in \cite{O}, the field strength propagator is given by
\be
\<H^{MNP}(x)H_{RST}(y)\> = \frac{c}{(s^2)^3}D^{MNP}{}_{RST}(I(s))
\ee
where $s=x-y$ for some normalization constant $c$. Definied in this way, it is easily seen that the bi-three-form $\<H(x)H(y)\>$ is invariant under inversions. We conclude that the gauge field propagator at most can change by at most a closed bi-two-form. 

In Feynman gauge and flat background we computed earlier the gauge field propagator to (dropping the normalization constant for the time being)
\be
\<B_{MN}(x)B^{PQ}(y)\> = \frac{\delta_{MN}^{PQ}}{|x-y|^4},
\ee
and we can check that it yields the field strength propagator above. Under inversion it transforms into 
\be
x^4 y^4 \frac{\delta_{MN}^{PQ}}{|x-y|^4}.
\ee
The bi-two-form propagator therefore transforms as
\bea
\<B(x)B(y)\>&\rightarrow& \frac{1}{(s^2)^2} I_{MM'}(x)I_{NN'}(x)I_{PM'}(y)I_{QN'}(y) dx^M\wedge dx^N(y) \otimes dy^P\wedge dy^{Q'}\cr
& = & \frac{1}{(s^2)^2} \(\delta_M^P\delta_N^Q-4\frac{\delta_M^Py_Ny^Q}{y^2}-4\frac{\delta_M^Px_Nx^M}{x^2}+8\frac{\delta_M^Px_Ny^Q(x\cdot y)}{x^2y^2}+8\frac{y_Mx_Ny^Px^Q}{x^2y^2}\)\cr
&&dx^M\wedge dx^N\otimes dy_P\wedge dy_Q.
\eea
The variation can be written as a bi-exact form,
\be
-2\delta_M^P\[\frac{\partial}{\partial x^N}\(\frac{y^Q}{|x-y|^2y^2}\)+\frac{\partial}{\partial y^Q}\(\frac{x_N}{|x-y|^2x^2}\)\]+4\frac{\partial}{\partial x^M}\(\frac{x_N x^P y^Q}{|x-y|^2x^2y^2}\).
\ee
just as we had expected, since $\<H(x)H(y)\>$ was invariant under inversion. Furthermore, since the gauge fixing term is invariant under all transformation except for inversions, we deduce from this result that for any conformal transformation which does not take us outside a flat space, the variation should be given by 
\be
\delta_M^P\[\frac{\partial}{\partial x^N}\(\frac{\partial^Q\phi(y)}{|x-y|^2}\)+\frac{\partial}{\partial y^Q}\(\frac{\partial_N\phi(x)}{|x-y|^2}\)\]+\frac{\partial}{\partial x^M}\(\frac{x_N\partial^P\phi(x) \partial^Q\phi(y)}{|x-y|^2}\)
\ee

The spherical Wilson surface of radius $R$ and which goes through the origin is given by
\be
W_0(S_R^2) = \frac{1}{\pi^2}\int_{S^2_{R,R}} d\tilde{X}^{M}\wedge d\tilde{X}^{N}\int_{S^2_{R,R+\epsilon}} d\tilde{Y}_{M}\wedge d\tilde{Y}_{N} \frac{1}{|\tilde{X}-\tilde{Y}|^4}
\ee 
Under inversion, $\tilde{X}^M=\frac{X^M}{|X|^2}$ the sphere transforms to a plane which has its closest distance to the origin equal to $a\equiv 1/(2R)$. That is, while $\tilde{X}^{M}$ take values on a sphere, $X^M$ take values on a plane $D_{\infty}$. We have
\be
W(S_R^2) = \frac{1}{\pi^2}\int_{D_{\infty}}dX^1\wedge dX^2\int_{?} dY^1\wedge dY^2 \(\frac{1}{|X-Y|^4}
+{\mbox{exact terms}}\)
\ee
Here $?$ is the complicated displaced surface along geodesics (in this case it will be a very big sphere). The exact terms would have been unimportant here, had it not been that the plane is infinitely extended. If we regularize by taking a big disk of radius $r$ instead of the infinite plane, we have to take into account the contribution of the exact terms coming from the boundary of the disk. This will have the effect of cancelling the divergence before letting the disk becoming infinitely big.
The contribution from the interior of the disk to the anomaly related to this transformation is given by
\be
{\cal{A}}=-\frac{1}{4\pi}\int_{D_r^2}d^2\sigma\(|\partial_{\alpha}\phi|^2+|\partial\phi|^2\)
\ee
for $\phi=-2\ln|X(\sigma)|$. Choosing coordinates 
\bea
X^1 &=& r\cos\varphi\cr
X^2 &=& r\sin\varphi
\eea
on the plane, we thus have to compute
\be
{\cal{A}}=-\frac{1}{\pi}\int dr r d\varphi\(\left|\frac{X_{\alpha}}{|X|^2}\right|^2+\left|\frac{X_M}{|X|^2}\right|^2\)
\ee
with $|X_{\alpha}|^2=r^2$ and $|X|^2=r^2+a^2$. We get
\be
{\cal{A}}=4\(-\ln (2rR) + \frac{1}{4}\)
\ee
but this is not the whole answer. We have to add the contribution coming from the variation of the propagator, i.e. from the boundary of the disk. Some computation shows that their contribution is $4\ln r$. We thus get the total anomaly as
\be
 {\cal{A}}_{TOTAL}=4\(-\ln (2R) + \frac{1}{4}\),
\ee
and indeed this is equal to the renormalized abelian spherical Wilson surface, consult Eq (\ref{sphere}).

If we consider the $A_1$ Wilson surface we instead get the anomaly
\be
-4\ln (2R).
\ee
Again we get the correct answer, that is, that we read off from our explicit computation given in Eq (\ref{sphere1}). This is a non-trivial check of Eq (\ref{anom}).

In the limit $N\rightarrow\infty$, we can just compute the contribution to the anomaly from the interior of the disk, which is
\be
-4N\ln (2rR)
\ee
since the form of the boundary term is not known, so we can not deduce from this what its contribution to the Weyl anomaly should be. But we know what the expectation value for the spherical Wilson surface is. This has been computed in \cite{BCFM} in AdS space, with the result
\be
4N\(-\ln(2R)-\frac{1}{2}\)
\ee
for large $N$. This should thus be equal to the conformal anomaly for the plane under inversion. We deduce that we have got an additional contribution of $-4N/2$ from the point at infinity. This contribution we think should come from some matrix model sitting at the point at infinity, in analogy with \cite{DG}.

We could of course had viewed all this in the opposite direction and computed the anomaly on a sphere instead of on a plane. We would then have got minus the results that we got above.

\section{Connecting to supersymmetric Yang-Mills theory}\label{ph}
In Yang-Mills theory the Wilson loop is derived from a correlation function $\<J(y)J(x)\>$ of quark-currents $J(x) \sim q(x)\bar{q}(x)$ that creates (annihilates) a quark-antiquark pair at x (y). The Wilson loop is what one gets if one drops the path integral over paths connecting $x$ and $y$, but just consider one such path \cite{W}. 

We are interested in defining a Wilson loop in $U(2)$ gauge theory, spontaneously broken to $U(1)\times U(1)$ by a large Higgs vacuum expectation value. Realizing this by two separated $D3$-branes, the quark fields will correspond to the end points of strings that end on the two parallel $D3$-branes. 

In Minkowski space with metric $\eta_{\mu\nu}=$diag$(-+++)$, the bosonic part of the $N=4$ supersymmetric action is
\be
-\frac{1}{4\pi}(\hat{F}_{\mu\nu},\hat{F}^{\mu\nu})-\frac{1}{2\pi}(\hat{D}_{\mu}\hat{\phi}_{A})^2+\frac{1}{4\pi}[\hat{\phi}_A,\hat{\phi}_B]^2
\ee
Following the steps given in the appendix in \cite{DGO}, we expand the scalar field around a Higgs VEV $\phi^{(0)}$ as
\be
\hat{\phi}_A = \phi^{(0)}_A + \phi
\ee
with
\bea
{\phi^{(0)}}_A &=& \delta_{A1}\frac{1}{2}\(\begin{array}{cc}
M & 0\\
0 & -M
\end{array}\)\cr
\phi &=& 
\frac{1}{2}\(\begin{array}{cc}
\varphi & w\\
w^{\dag} & -\varphi
\end{array}\)
\eea
and decompose the gauge potential as
\bea
\hat{A}_{\mu}=\frac{1}{2}\(\begin{array}{cc}
A_{\mu} & W_{\mu}\\
W^{\dag}_{\mu} & -A_{\mu}
\end{array}\)
\eea
and insert this in the action. We then consider the correlation function
\be
\<w(x)^{\dag}w(x)w(y)^{\dag}w(y)\>.
\ee
We then will find that the Wilson loop should be defined as
\be
\exp \(i\int ds\(A_{\mu}\dot{X}^{\mu}-\varphi\sqrt{|\eta_{\mu\nu}\dot{X}^{\mu}\dot{X}^{\nu}}|\)\),
\ee 
if $s$ parametrizes the loop $s\mapsto X^{\mu}(s)$. Wick rotating amounts to the replacement $X^0\rightarrow iX^0$, $A_0\rightarrow -iA_0$. If $-\eta_{\mu\nu}\dot{X}^{\mu}\dot{X}^{\nu}>0$, then the euclidean version should read
\be
\exp \(i\int ds\(A_{\mu}\dot{X}^{\mu}-i\varphi\sqrt{\delta_{\mu\nu}\dot{X}^{\mu}\dot{X}^{\nu}}\)\),
\ee 
but if $-\eta_{\mu\nu}\dot{X}^{\mu}\dot{X}^{\nu}<0$ we instead get
\be
\exp \(i\int ds\(A_{\mu}\dot{X}^{\mu}-\varphi\sqrt{\delta_{\mu\nu}\dot{X}^{\mu}\dot{X}^{\nu}}\)\).\label{loop}
\ee 
It is really this latter situation that we study in this letter. We do not understand how a generic closed loop in Minkowski space  extended in the time-direction is to be interpreted in euclidean space. This problem is avoided by considering only closed spatial loops at a fixed time since such loops exist in both signatures.

Compactifying the six-dimensional theory on a two-torus, and letting the Wilson surface be a cylinder wrapping one of the compact dimensions, we find that $W_1$ defined in Eq (\ref{Wilson1}) reduces to (\ref{loop}). (If the surface instead wraps the other cycle it reduces to a 't Hooft loop.)

We can not reduce a six-dimensional action to get the four-dimensional effective action because the gauge field that reduces to the four-dimensional gauge field is selfdual, so its 6d action is zero. But we can reduce the 6d action for the scalar field. We can also reduce the $(2,0)$ supersymmetry transformations from 6d to 4d. This gives us sufficiently much information to construct the dimensionally reduced theory and in particular derive the normalization for the kinetic term for the 4d gauge field from this supersymmetry. 

Now $(2,0)$ supersymmetry is best understood in Minskowski space with metric $G_{MN}=(-1,1,1,1,1,1)$ because the spinor $\psi$ in the tensor multiplet is a symplectic Majorana spinor that only can be definied in a invariant way in Minskowski signature. Defining  $\bar{\psi}\equiv \psi^{\dag}\Gamma^0$ it is then constrained by $\bar{\psi}=\psi^Tc\Omega$ where $c$ and $\Omega$ are charge conjugation matrices, and $\Gamma_M$ and $\sigma_a$ denote intertwining matrices for $SO(1,5)$ and $SO(5)_R$ respectively, obeying $(\Gamma^M)^T=-c\Gamma^M c^{-1}$ and $(\sigma^a)^T=\Omega\sigma^a\Omega^{-1}$. The following $(2,0)$ supersymmetry transformations 
\bea
\delta B_{MN}&=&i\bar\epsilon \Gamma_{MM}\psi\cr
\delta \phi^a&=&i\bar\epsilon \sigma^a\psi\cr
\delta \psi&=&\(\frac{1}{12}\Gamma^{MNP}H_{MNP}-\Gamma^M\sigma_a\partial_M\phi^a\)\epsilon\cr
\delta\bar{\psi}&=&\bar{\epsilon}\(\frac{1}{12}\Gamma^{MNP}H_{MNP}+\Gamma^M\sigma_a\partial_M\phi^a\)
\eea
leave the following action
\be
\frac{1}{4\pi}\int\(-\frac{1}{2}H\wedge*H-d\phi^a\wedge *d\phi^a + i\bar{\psi}\Gamma^M\partial_M\psi\)\label{six}
\ee
invariant. Here $H$ is non-selfdual. Letting $\epsilon$ become space-time dependent we get the corresponding Noether supercharges as\footnote{This approach was also taken in \cite{AFH}, though in a different notation and conventions.}
\be
-4\pi Q=\int d^5x \(\frac{1}{6}\Gamma^{MNP}\Gamma^0H_{MNP}+2\Gamma^M\Gamma^0\sigma_a\partial_M\phi^a\)\psi.
\ee
We make the decomposition ($\sigma^{1,2,3}$ are Pauli matrices, not be confused with $\sigma^a$)
\bea
\Gamma^{\mu} & = & \gamma^{\mu} \otimes 1\otimes 1\cr
\Gamma^4 & = & \gamma \otimes \sigma^1\otimes 1\cr
\Gamma^5 & = & \gamma \otimes \sigma^2\otimes 1\cr
\Sigma^a &=& 1\otimes\sigma^1 \otimes \sigma^a\cr
\Sigma^6 &=& 1\otimes\sigma^2 \otimes 1,
\eea
and the following field redefinitions ($R$ denote the radii of a rectangular two-torus)
\bea
RB_{\mu 4} &=& A_{\mu}\cr
RB_{\mu 5} &=& \tilde{A}_{\mu}\cr
RB_{45} &=& \Phi^6\cr
R \phi^a&=&\Phi^a\cr
R \psi&=&\gamma_0\gamma\otimes \sigma^1\otimes 1\Psi.
\eea
Dimensional reduction amounts to disregarding derivatives in compact dimensions. Letting $A=(a,6)$, the supercharges reduce to
\be
-4\pi Q=\int d^3 x\(\gamma^{\mu\nu}F_{\mu\nu}+2\gamma^{\mu}\gamma\Sigma_A\partial_{\mu}\Phi^A\)\Psi.
\ee
Now $A$ will be a vector index in the enlarged R-symmetry group $SO(6)_R$. This is to be compared to the supercharges in abelian $N=4$ supersymmetric gauge theory \cite{B}. From these reduced supercharges we deduce that the theory that one gets from the chiral part of (\ref{six}) is given by the action 
\be
\frac{1}{4\pi}\(-F\wedge*F-d\phi\wedge*d\phi+i\Psi\gamma^{\mu}\partial_{\mu}\Psi\).
\ee
We notice that we get the normalization that one should have in order to have $\int F/(2\pi)$ integer quantized. 

We should however notice that the Wilson surface is subject to a periodicity condition,
\be
\int_{T^2} dx^4 dx^5 B_{45} \sim \int_{T^2} dx^4 dx^5 B_{45} + 2\pi \bf{Z}
\ee
which implies that $B_{45}$ is only defined modulo $\frac{2\pi}{R^2}$ which in turn implies that 
\be
\Phi^6 \sim \Phi^6 + \frac{2\pi}{R}\bf{Z},
\ee
so in the compactification limit $R\rightarrow 0$ this compact scalar decompactifies.

\vskip 0.5truecm

\noindent{\sl{Acknowledgements}}:

\noindent{I am grateful to Lars Brink for having encouraged me, and to Mans Henningson for having answered some questions.}


\vskip 0.5truecm



\newpage

\begin{thebibliography}{999}
\bibitem{Ga}O. Ganor, `Six-dimensional tensionless strings in the large N limit', Nucl. Phys. B489:95-121 (1997), hep-th/9605201.
\bibitem{GW}R. Graham, E. Witten, `Conformal anomaly of submanifold observables in AdS/CFT correspondence', Nucl. Phys. B546:52-64 (1999), hep-th/9901021. 
\bibitem{BCFM}D. Berenstein, R. Corrado, W. Fischler, J. Maldacena, `The operator product expansion for Wilson loops and surfaces in the large N limit', Phys.  Rev.D59:105023 (1999) hep-th/9809188.
\bibitem{DG}N. Drukker, D. Gross, `An exact prediction of N=4 SUSYM theory for string theory', J. Math. Phys. 42:2896-2914 (2001), hep-th/0010274.
\bibitem{HS}M. Henningson, S. Skenderis `Weyl anomaly for Wilson surfaces', JHEP 9906:012 (1999), hep-th/9905163. 
\bibitem{Gu}A. Gustavsson, `On the Weyl anomaly of Wilson surfaces', JHEP 0312:059 (2003), hep-th/0310037.
\bibitem{O}H. Osborn, `Implications of conformal invariance for quantum field theories in $d>2$', Ahrenshoop Symp.1993:0014-26, hep-th/9312176.
\bibitem{W}K. G. Wilson, `Confinement of quarks', Phys. Rev. D:2445-2459 (1974).
\bibitem{DGO}N. Drukker, D. J. Gross, H. Ooguri, `Wilson loops and minimal surfaces', Phys.Rev.D60:125006 (1999), hep-th/9904191.
\bibitem{AFH}P. Arvidsson, E. Flink, M. Henningson, `Free tensor multiplets and strings in spontaneously broken six-dimensional $(2,0)$ theory', JHEP 0306:039 (2003), hep-th/0306145.
\bibitem{B}L. Brink, J. H. Schwarz, J. Scherk, Nucl. Phys. B121 (1977) 77.
\bibitem{Henning}M. Henningson, `The quantum Hilbert space of a chiral two-form in d=5+1 dimensions', JHEP 0203, 021 (2002), hep-th/0111150.
\bibitem{Gustav}A. Gustavsson, `The d=6, (2,0)-tensor multiplet coupled to self-dual strings', Int. J. Mod. Phys. A17 (2002) 2051-2072, hep-th/0110248.
\end{thebibliography}
\end{document}